\documentclass[%
twocolumn,
 amsmath,amssymb,
 aps,
pra,
showpacs
]{revtex4-1}
\usepackage[T1]{fontenc}
\usepackage[latin9]{inputenc}
\usepackage{geometry}
\usepackage{float}
\usepackage{amsmath}
\usepackage{amssymb}
\usepackage{graphicx}
\usepackage{epstopdf}



\begin{document}

\title{Exact analysis of a Veselago lens in the quasi-static regime}
\author{Asaf Farhi}
\email{asaffarhi@post.tau.ac.il}
\author{David J. Bergman}
\email{bergman@post.tau.ac.il}
\affiliation{
Raymond and Beverly Sackler School of Physics and Astronomy,
Faculty of Exact Sciences,
Tel Aviv University, IL-6997801 Tel Aviv, Israel
}
\date{\today}
\begin{abstract}
The resolution of conventional optical lenses is limited by the wavelength.
Materials with negative refractive index have been shown to enable
the generation of an enhanced resolution image where both propagating
and non-propagating waves are employed. We analyze such a Veselago
lens by exploiting some exact one dimensional integral expressions
for the quasi-static electric potential of a point charge in that
system. Those were recently obtained by expanding that potential in
the quasi-static eigenfunctions of a three-flat-slabs composite structure.
Numerical evaluations of those integrals, using realistic values for
physical parameters like the electric permittivities of the constituent
slabs and their thickness, reveal some surprising effects: E.g.,
the maximum concentration of the electric field occurs not at the geometric
optics foci but at the interfaces between the negative permittivity
slab and the positive permittivity slabs. The analysis provides simple
computational guides for designing such structures in order to achieve
enhanced resolution of an optical image.
\end{abstract}

\pacs{78.20.Bh, 42.79.-e, 42.70.-a}

\maketitle

\section{Introduction}
The resolution limit in conventional optical imaging is known to be inversely proportional to the wavelength of the light. In 1967 a theoretical analysis by Veselago, based upon geometric optics, suggested that a flat slab with a negative refractive index can focus at a point the radiation from a point source \cite{veselago1967electrodinamics}.
At that time materials possessing negative refractive index did not exist but recent developments in metamaterials have made the production of such materials possible \cite{sievenpiper19963d,pendry1996extremely,pendry1999magnetism}.
In 2000 another important analysis by Pendry showed that materials with a negative refractive index can amplify evanescent waves, and thus enable the generation of an image by both propagating and non propagating waves, theoretically leading to unlimited resolution \cite{pendry2000negative}.     

In the quasistatic regime, when the typical length scales are much smaller than the wavelength, Maxwell's equations reduce to static equations in which the electric and the magnetic fields are decoupled. 
Hence, the optical constant of relevance in this regime is the electric permittivity rather than the refractive index.
The imaging of an electric point charge was recently analyzed by expanding the local electric potential in a series of the quasi-static eigenfunctions of a three-flat-slabs composite structure.  This analysis yielded exact 
one dimensional integral expressions for the quasi-static electric potential of a point charge in that
system \cite{bergman2014perfect,Bergman2013static}. 

In this communication we first derive exact expressions for the electric field in a two constituent three-flat-slabs composite structure in the form of one dimensional integrals. We then perform numerical computations for such a setup using realistic values for the 
physical parameters like the electric permittivities and the thickness of the intermediate slab. In these computations we vary both the location of the point charge and the constituent permittivities of the medium. These computations reveal surprising results among which is that the best imaging is obtained at the interfaces between the intermediate slab and the surrounding medium rather than at the geometric optics foci.     
  
The structure of the paper is as follows. In Section \ref{section:summary_theory} we present a summary of the basic theory for the analysis of such a setup. In Section \ref{section:expressions_electric_field} we derive exact expressions for the local electric field and validate our results. In Section \ref{section:numerical_computation} we present results of the numerical computations for various charge locations and for various permittivity values. In Section \ref{section:discussion} we discuss our results. 
          
\section{Summary of the basic theory}
\label{section:summary_theory}
\begin{figure}
\centerline{
\includegraphics[width= 8.0cm]{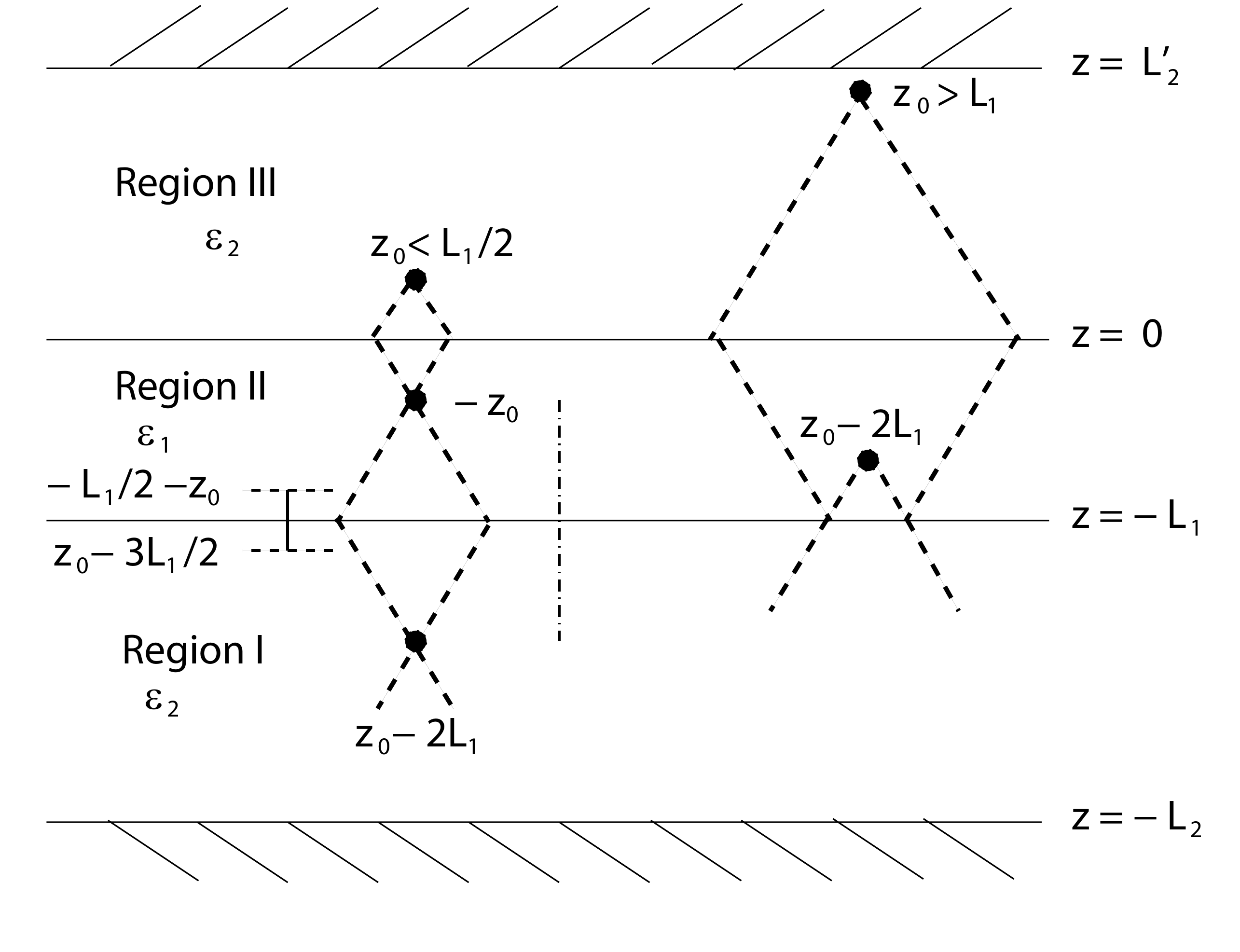}}
\caption{A three-parallel-slabs microstructure that fills the entire volume of a large parallel-plate
capacitor. The upper layer (Region III), where $\epsilon=\epsilon_2$, includes a point charge $q$ located at
${\bf r}_0=(0,0,z_0)$. In the left part $z_0<L_1/2$ while in the right part $z_0>L_1$,
where $L_1$ is the thickness of the intermediate $\epsilon_1$ layer (Region II).
 Even when all the other linear sizes of this structure tend to $\infty$,
this configuration is still unsolvable in any simple fashion. The diagonal dashed lines show how a geometric
optics or light rays description would lead to a focusing of the original point charge in Region III at new points in Regions I
and II when $\epsilon_2=-\epsilon_1$.
The vertical dot-dashed line indicates the regions where $\psi({\bf r})$ then diverges in the case
shown on the left side, while the vertical solid line shows where the dissipation rate diverges (after reference \cite{bergman2014perfect}).
%\vspace{-5 true mm}
}
\label{ThreeSlabsMicrostructureA}
\end{figure}

In this section we describe the derivation of the exact results for the local electric potential field $\psi({\bf r})$ in the quasistatic limit for the case of a point electric
charge $q$ in a two-constituent composite medium \cite{bergman2014perfect,Bergman2013static}. In these references a generic two-constituent composite structure, composed of three infinitely wide parallel slabs, is considered. The intermediate slab, with an electric permittivity $\epsilon_1$, is placed between two slabs with  an electric permittivity $\epsilon_2$ (see Fig. \ref{ThreeSlabsMicrostructureA}).

In the static limit Maxwell's equations reduce to Poisson's equation for $\psi({\bf r})$:
\begin{equation}
\label{PhiDiffEq}
 -4\pi\rho({\bf r})=\nabla\cdot(\epsilon_1\theta_1+\epsilon_2\theta_2)\nabla\psi,
\end{equation} 
which can be rewritten as:
\begin{equation}
\label{PhiDiffEq2}
\nabla^{2}\psi=-4\pi\rho\left(\mathbf{r}\right)/\epsilon_{2}+u\nabla\cdot\left(\theta_{1}\nabla\psi\right),
\end{equation} 
\begin{equation*}
\theta_{1}({\bf r})\equiv1-\theta_{2}({\bf r})=\left\{ \begin{array}{lcl}
1 & {\rm if} & \epsilon({\bf r})=\epsilon_{1}\\
0 & {\rm if} & \epsilon({\bf r})=\epsilon_{2}
\end{array}\right\} ,\, u\equiv1-\frac{\epsilon_{1}}{\epsilon_{2}}, 
\end{equation*}
where $\theta_1$ and $\theta_2\equiv 1-\theta_1$ are step functions that characterize the microstructure
of the composite medium. The function $\rho({\bf r})$ which appears in these equations 
represents a charge density distribution, including the possibility that $\rho({\bf r})=q\delta^3({\bf r}-{\bf r}_0)$,
i.e., a point charge at ${\bf r}_0$. The capacitor plates at $z=-L_2$ and $z=L'_2$ are included in
order that appropriate boundary conditions may be imposed there so as to result in a unique
solution for $\psi({\bf r})$.  At the end of the calculation we will take the limits $L_2\rightarrow\infty$ and $L'_2\rightarrow\infty$.

We reformulate Eq. (\ref{PhiDiffEq2}) as an  integro-differential equation for
$\psi({\bf r})$ \cite{LectNotes83}:
\begin{eqnarray}
\psi({\bf r})&=&\psi_0({\bf r})+u\hat\Gamma\psi,\label{IntegroDiffEq}\\
\hat\Gamma\psi&\equiv&\int  dV' \theta_1({\bf r}')\nabla' G_0({\bf r},{\bf r}')\cdot \nabla'  \psi({\bf r}'),\nonumber
\label{GammaDef}
\end{eqnarray}
where $G_0({\bf r},{\bf r}')$  is Green's function for Laplace's equation with zero boundary conditions defined as follows:
\begin{eqnarray*}
\nabla^2G_0({\bf r},{\bf r}')&=&-\delta^3({\bf r}-{\bf r}),%\nunumber
\label{GreenDiffEq}\\
G_0({\bf r},{\bf r}')&=&0 \;\; {\rm for}\; z=-L_2\; {\rm and}\; z=L'_2,%\nunumber
\label{GreenBoundaryCond}
\end{eqnarray*}
and $\psi_0({\bf r})$ is the solution of Poisson's equation in a uniform medium with a permittivity $\epsilon_2$.
 
In the case of no charges and vanishing boundary conditions, Eq (\ref{PhiDiffEq}) reduces to 
\begin{equation*}
s\psi({\bf r})=\hat\Gamma\psi,\label{IntegroDiffEqHomegeous}\, s\equiv\frac{1}{u}.
\end{equation*}
Defining the scalar product of two scalar functions $\psi(\bf r),\phi(\bf r)$ by
$$\left\langle \psi|\phi\right\rangle \equiv \int d^{3}r\theta_{1}\nabla\psi^{*}\cdot\nabla\phi $$
makes $\hat\Gamma$ a Hermitian operator \cite{LectNotes83}. Therefore it has a complete set of  eigenfunctions $\phi_n$ and eigenvalues $s_n$
\begin{equation*}
s_n\phi_n({\bf r})=\hat\Gamma\phi_n.
\label{GammaEigenstates}
\end{equation*}
By using the expansion of the unity operator $\hat I$ in Eq. (\ref{IntegroDiffEq}),  we can expand the potential in a series of the eigenfunctions $\phi_n$:
\begin{eqnarray}
\hat I&=&\sum_n |\phi_n\rangle\langle \phi_n|\notag
\label{UnityOperator}\\
&&\hspace{-3 mm}\Longrightarrow
 \psi({\bf  r})=\psi_0({\bf  r})+\sum_n \frac{s_n}{s-s_n}\langle\phi_n|\psi_0\rangle \phi_n({\bf  r}).
\label{PhiExpansion}
\end{eqnarray}
We now set the charge distribution to be that of a point charge located at $\mathbf{r}_{0}=\left(0,0,z_{0}\right)$. This means that  
\begin{equation}
\psi_0({\bf r})=\frac{q/\epsilon_2}{|{\bf r}-{\bf r}_0|}.
\label{CoulombPot}
\end{equation}
The eigenfunctions that satisfy Laplace's equation with vanishing boundary conditions are:
\begin{eqnarray}
\lefteqn{\phi^\pm_{\bf k}({\bf r})=e^{i{\bf k}\cdot{\mathbf\rho}}\cdot}\nonumber\\
&&\hspace{-8 true mm}\cdot \left\{\begin{array}{ll}
A^\pm_{\bf k} \sinh[k(z+L_2)], & z\in {\rm I},\\
B^\pm_{\bf k}\sinh(kz) + B'^\pm_{\bf k}\sinh[k(z+L_1)], &  z\in {\rm II},\\
C^\pm_{\bf k}\sinh[k(z-L'_2)], &  z\in {\rm III}.
\end{array}\right.
\label{ThreeSlabsEigenstates}
\end{eqnarray}
By imposing continuity of the potential and the perpendicular component of $\mathbf{D}$ and taking the limits $L'_2,L_2 \rightarrow\infty$, we get the eigenvalues and the coefficients in these expressions
\begin{eqnarray*}
 s^\pm_{\bf k}= \frac{1\mp e^{-kL_1}}{2}\,,A^\pm_{\bf k}=-B^\pm_{\bf k}\frac{\sinh(kL_1)}{\sinh[k(L_2-L_1)]},\\
B'^\pm_{\bf k}=\mp B^\pm_{\bf k}\,,C^\pm_{\bf k}=\pm B^\pm_{\bf k}\frac{\sinh(kL_1)}{\sinh(kL'_2)}.\,\,\,\,\,\,\,\,\,\,\,\,\,\,\,\,\,\,\,
\end{eqnarray*}
The normalization condition $\langle\phi^\pm_{\bf k}|\phi^\pm_{\bf k}\rangle=1$ leads to 
$$1=2k L_x L_y |B_{\bf k}^\pm|^2\sinh(kL_1)\left[\cosh(kL_1)\mp 1\right].$$
 The eigenvalues have a single accumulation point at $s=1/2$  which is therefore a very singular point of Eq. (\ref{PhiExpansion}).
That equation leads to the following expressions for the electric potential in the three regions in the form of one dimensional integrals \cite{bergman2014perfect}:
\begin{widetext}
\begin{eqnarray}
\hspace{-2 true mm}\psi\hspace{-1 true mm}&\hspace{-1 true mm}=&\hspace{-1 true mm}\frac{4s(1-s)q}{\epsilon_2}\int_0^\infty dk\, J_0(k\rho)\,
 \frac{e^{-k(z_0-z)}}{e^{-2kL_1}-(2s-1)^2}
= 4q\epsilon_1\int_0^\infty dk\, J_0(k\rho)\,
 \frac{e^{-k(z_0-z)}}{(\epsilon_2-\epsilon_1)^2e^{-2kL_1}-(\epsilon_2+\epsilon_1)^2}
 \; \mbox{in I},\label{psi1}\\
\psi&\hspace{-2 true mm}=&\frac{2sq}{\epsilon_2}\int_0^\infty dk\, J_0(k\rho)\,e^{-k(z_0-z)}
 \frac{e^{-2k(z+L_1)}-2s+1}{e^{-2kL_1}-(2s-1)^2}\nonumber\\
&=&2q\int_0^\infty dk\, J_0(k\rho)\,e^{-k(z_0-z)}\,
 \frac{(\epsilon_2-\epsilon_1)e^{-2k(z+L_1)}-(\epsilon_2+\epsilon_1)}
  {(\epsilon_2-\epsilon_1)^2e^{-2kL_1}-(\epsilon_2+\epsilon_1)^2}
 \; \mbox{in II},\label{psi2}\\
\psi&=&\psi_0+\frac{q(2s-1)/\epsilon_2}{\sqrt{\rho^2+(z+z_0)^2}}
 -\frac{4s(1-s)(2s-1)q}{\epsilon_2}\int_0^\infty dk\, J_0(k\rho)\,
\frac{e^{-k(z_0+z)}}{e^{-2kL_1}-(2s-1)^2}
=\frac{q/\epsilon_2}{\sqrt{\rho^2+(z-z_0)^2}}\nonumber\\
&& +\;\frac{\epsilon_2+\epsilon_1}{\epsilon_2-\epsilon_1}
  \frac{q/\epsilon_2}{\sqrt{\rho^2+(z+z_0)^2}}
+4q\epsilon_1\frac{\epsilon_2+\epsilon_1}{\epsilon_2-\epsilon_1}
 \int_0^\infty dk\, J_0(k\rho)\,
\frac{e^{-k(z_0+z)}}{(\epsilon_2-\epsilon_1)^2e^{-2kL_1}-(\epsilon_2+\epsilon_1)^2}
 \; \mbox{in III}.\label{psi3}
\end{eqnarray}
\end{widetext}
These expressions for the potential, as well as the local dissipation rate, defined by Im$\left(\epsilon\right)\left|\mathbf{E}\right|^{2}/8\pi$, were analyzed for the case  of $s=1/2$ (i.e., $\varepsilon_{1}=-\varepsilon_{2}$) \cite{bergman2014perfect}. This analysis showed that the potential diverges  in the range of positions $z_0-2L_1<z<-z_0$. Moreover, when the location of the point charge satisfies $z_0<L_1/2$, the local dissipation rate diverges for $z$ in the range $\left[z_{0}-3L_{1}/2,-z_{0}-L_{1}/2\right]$ (see Fig. \ref{ThreeSlabsMicrostructureA}).

When $s=1/2$ these expressions for the potential take the following exact
closed forms in those regions of $z$ where it is non-diverging:
\vspace{-2 true mm}
$$
\vspace{-2 true mm}
\psi({\bf r})=\left\{\begin{array}{ll}
\frac{q/\epsilon_2}{\sqrt{\rho^2+(z-z_0+2L_1)^2}}, & {\bf r}\in {\rm I},\\
\frac{q/\epsilon_2}{\sqrt{\rho^2+(z+z_0)^2}}, & {\bf r}\in {\rm II},\\
\frac{q/\epsilon_2}{\sqrt{\rho^2+(z-z_0)^2}}, & {\bf r}\in {\rm III}.
\end{array}\right.$$
%\vspace{-2 true mm}
This means that the potential above the top geometric optics image at $\mathbf{r}=(0,0,-z_0)$ and below the bottom geometric optics image at $\mathbf{r}=(0,0,z_0-2L_1)$, is that of a point charge located at these points. In the intermediate $z$ values between these expected images, the potential diverges (see Fig. \ref{ThreeSlabsMicrostructureA}). Since there are no point charges located at these points, the surface integration over the electric field perpendicular to an arbitrary envelope surrounding one of these points gives zero according to Gauss' law. This is fulfilled since the contribution to the surface integral from where the potential diverges cancels out with the contribution from where the potential is finite (for a spherical surface centered around one of these points, the first and second contributions give $-q/2$ and $q/2$ respectively).
\section{Exact expressions for the electric field and verification of the results}
\label{section:expressions_electric_field}
We calculated exact expressions for the electric fields by differentiating
the expressions for the potentials derived in \cite{bergman2014perfect} and reproduced in Eqs. (\ref{psi1})-(\ref{psi3}) with respect to $\rho$ and $z$.
The expressions for the $z$ and $\rho$ components of $\mathbf{E}$ are as follows, where we substituted $s\equiv 1/2+\Delta s$:
\subsection*{Region I}
\begin{eqnarray}
E_{\textrm{I}\,\rho}=C_{1}\intop_{0}^{\infty}dkkJ_{1}\left(k\rho\right)\frac{e^{-k\left(z_{0}-z\right)}}{e^{-2kL_{1}}-4\left(\Delta s\right)^{2}},\\
E_{\textrm{I}\, z}=-C_{1}\intop_{0}^{\infty}dkkJ_{0}\left(k\rho\right)\frac{e^{-k\left(z_{0}-z\right)}}{e^{-2kL_{1}}-4\left(\Delta s\right)^{2}}, 
\end{eqnarray}
%\begin{equation}
%%E_{\textrm{I}_{\rho}}\left(\mathbf{r}\right)=\frac{4q\left(\frac{1}{2}+\Delta s\right)\left(\frac{1}{2}-\Delta s\right)}{\epsilon_{2}}\intop_{0}^{\infty}dkkJ_{1}\left(k\rho\right)\frac{e^{-k\left(z_{0}-z\right)}}{e^{-2kL_{1}}-4\left(\Delta s\right)^{2}}
%E_{\textrm{I}\, \rho}=C_{1}\intop_{0}^{\infty}dkkJ_{1}\left(k\rho\right)\frac{e^{-k\left(z_{0}-z\right)}}{e^{-2kL_{1}}-4\left(\Delta s\right)^{2}},
%\end{equation}
%\begin{equation}
%%E_{\textrm{I}_{z}}\left(\mathbf{r}\right)=-\frac{4q\left(\frac{1}{2}+\Delta s\right)\left(\frac{1}{2}-\Delta s\right)}{\epsilon_{2}}\intop_{0}^{\infty}dkkJ_{0}\left(k\rho\right)\frac{e^{-k\left(z_{0}-z\right)}}{e^{-2kL_{1}}-4\left(\Delta s\right)^{2}}
%E_{\textrm{I}\, z}=-C_{1}\intop_{0}^{\infty}dkkJ_{0}\left(k\rho\right)\frac{e^{-k\left(z_{0}-z\right)}}{e^{-2kL_{1}}-4\left(\Delta s\right)^{2}},
%\end{equation}
where
\begin{equation*}
C_{1}\equiv\frac{q\left[1-4\left(\Delta s\right)^{2}\right]}{\epsilon_{2}}.
\end{equation*}
\subsection*{Region II}
\begin{eqnarray}
E_{\textrm{II}\, \rho}=C_{2}\intop_{0}^{\infty}dkkJ_{1}\left(k\rho\right)e^{k\left(z-z_{0}\right)}\frac{e^{-2k\left(z+L_{1}\right)}-2\Delta s}{e^{-2kL_{1}}-4\left(\Delta s\right)^{2}}\label{eq:er_reg_2},\\
%\end{equation}
%\begin{equation}
E_{\textrm{II}\, z}=C_{2}\intop_{0}^{\infty}dkkJ_{0}\left(k\rho\right)e^{k\left(z-z_{0}\right)}\frac{e^{-2k\left(z+L_{1}\right)}+2\Delta s}{e^{-2kL_{1}}-4\left(\Delta s\right)^{2}}\label{eq:e_z_reg_2},
\end{eqnarray}
where
\begin{equation*}
C_{2}\equiv\frac{\left(1+2\Delta s\right)q}{\epsilon_{2}}.
\end{equation*}
\subsection*{Region III}
\begin{widetext}\begin{eqnarray}
E_{\textrm{III}\, \rho}=\frac{q}{\epsilon_{2}}\frac{\rho}{\left[\rho^{2}+\left(z-z_{0}\right)^{2}\right]^{3/2}}+\frac{2q\Delta s}{\epsilon_{2}}\frac{\rho}{\left[\rho^{2}+\left(z+z_{0}\right)^{2}\right]^{3/2}}-2C_{1}\Delta s\intop_{0}^{\infty}dkkJ_{1}\left(k\rho\right)\frac{e^{-k\left(z_{0}+z\right)}}{e^{-2kL_{1}}-4\left(\Delta s\right)^{2}},\\
%\end{equation}
%$$E_{\textrm{III}_{z}}=\frac{q}{\epsilon_{2}}\left(z-z_{0}\right)\left(\rho^{2}+\left(z-z_{0}\right)^{2}\right)^{-3/2}+\frac{2q\Delta s}{\epsilon_{2}}\left(z+z_{0}\right)\left(\rho^{2}+\left(z+z_{0}\right)^{2}\right)^{-3/2}$$
%\begin{equation}
%-2C_{1}\Delta s\intop_{0}^{\infty}dkkJ_{0}\left(k\rho\right)\frac{e^{-k\left(z_{0}+z\right)}}{e^{-2kL_{1}}-4\left(\Delta s\right)^{2}}
E_{\textrm{III}\, z}=\frac{q}{\epsilon_{2}}\frac{\left(z-z_{0}\right)}{\left[\rho^{2}+\left(z-z_{0}\right)^{2}\right]^{3/2}}+\frac{2q\Delta s}{\epsilon_{2}}\frac{\left(z+z_{0}\right)}{\left[\rho^{2}+\left(z+z_{0}\right)^{2}\right]^{3/2}}-2C_{1}\Delta s\intop_{0}^{\infty}dkkJ_{0}\left(k\rho\right)\frac{e^{-k\left(z_{0}+z\right)}}{e^{-2kL_{1}}-4\left(\Delta s\right)^{2}}.
\end{eqnarray}
\end{widetext}
In order to verify the expressions for the potential and the electric
field we checked the continuity of the
potential and the perpendicular component of $\mathbf{D}$ at the interfaces.
This was done by substituting $z=-L_{1}$ in the expressions for Regions I and
II and $z=0$ in the expressions for Regions II and III,
yielding the same expressions in both cases (see the Appendix for more details).
\section{Numerical Computations}
\label{section:numerical_computation}
We computed the one dimensional integrals in the expressions for the
potential and the electric field using Matlab. We verified the computations
of these integrals by checking the continuity of the potential and the
perpendicular component of $\mathbf{D}$ at the interfaces for a set of $\rho$
values (numerical values were compared).
In addition, we calculated the field intensity $I\left(\mathbf{r}\right)$ as well as the
dissipation rate $W\left(\mathbf{r}\right)$ in the 3 regions using the following definitions:
\begin{eqnarray}
I\left(\mathbf{r}\right)&\equiv&\left|E_{\rho}\left(\mathbf{r}\right)\right|^{2}+\left|E_{z}\left(\mathbf{r}\right)\right|^{2},\\
W\left(\mathbf{r}\right)&\equiv&\mathrm{Im}\left[\epsilon\left(\mathbf{r}\right)\right]\left(\left|E_{\rho}\left(\mathbf{r}\right)\right|^{2}+\left|E_{z}\left(\mathbf{r}\right)\right|^{2}\right).
\end{eqnarray}
%\begin{equation}
%I\left(\mathbf{r}\right)=\frac{c}{8\pi}\left(\left|E_{\rho}\left(\mathbf{r}\right)\right|^{2}+\left|E_{z}\left(\mathbf{r}\right)\right|^{2}\right),
%\end{equation}
%\begin{equation}
%W\left(\mathbf{r}\right)=\frac{\omega}{2}\mathrm{Im}\left[\epsilon\left(\mathbf{r}\right)\right]\left(\left|E_{\rho}\left(\mathbf{r}\right)\right|^{2}+\left|E_{z}\left(\mathbf{r}\right)\right|^{2}\right).
%\end{equation}
These are in fact the expressions for the intensity and the dissipation in which $c/8\pi$ and $\omega/8\pi$ were not included, respectively, for simplicity. 

We then placed another charge horizontally shifted from the original
charge in order to find the charge separation that is needed for resolution 
of the images. We varied that separation until the field intensity at the midpoint between the two images was $1/e^{1/2}$
of the intensity at the images. We defined this distance as the separation
distance needed to resolve the two images. In order to estimate
the resolution in each horizontal layer we normalized the local intensity in Region I by dividing it by the intensity at the horizontal coordinates of the images in that layer (see Figs. \ref{fig_two_charges_z0_8_7_L1}, \ref{fig_2_charges_z0_3_4_L1}, \ref{fig_2_charges_z0_3_8_L1}, \ref{fig_2_charges_z0_3_4_L1_imag_ds_div_100}, \ref{fig_2_charges_z0_3_4_L1_real_imag_ds_div_100}).

Throughout the computations we used $q=e$, where $e$ is the electron charge. We present the results for $\psi,\,I$ and $W$ without specifying units. Thus, in order for those results to be in 
 units of statV, erg/$\left(\mathrm{s}\cdot\mathrm{cm}^{2}\right)$ and erg/$\left(\mathrm{s}\cdot\mathrm{cm}^{3}\right)$, one has to multiply them by $q/e,q^2c/ 8\pi e^2$ and $q^2\omega/8\pi e^2$ respectively. 
\subsection{PMMA-silver-photoresist setup for different vertical charge locations}
\label{subs:different_vertical_charge_locations}
We modeled a PMMA-silver-photoresist setup that is similar to the one used
in \citep{liu2007far} by a two constituents setup in which the two
external slabs have the average permittivity value of PMMA and the photoresist,
and the permittivity of the intermediate slab is that of metallic silver. We used
the values for the permittivities suitable for a free space wavelength of $365$nm
\citep{liu2007far} :
\begin{eqnarray*}
\epsilon_{\mathrm{silver}}=-2.55+0.24i,\\
\epsilon_{\mathrm{PMMA}}=2.25+0.12i,\\
\epsilon_{\mathrm{PR}}=2.886+0.059i,
\end{eqnarray*}
%\[\epsilon_{\mathrm{silver}}=-2.55+0.24i,\]
%\[
%\epsilon_{\mathrm{PMMA}}=2.25+0.12i,
%\]
%\[
%\epsilon_{\mathrm{PR}}=2.886+0.059i,
%\]
which lead in the two constituents setup to the following permittivity values:
\[
\epsilon_{\mathrm{1}}=-2.55+0.24i,\,\epsilon_{2}=2.57+0.0896i.
\]
The silver slab thickness was set to $L_{1}=35$nm as in \citep{liu2007far}
and the external slabs in the calculation are assumed to
have infinite thickness.
We performed the computations for several locations of the point charge object
on the vertical axis. The first location was $z_{0}=40$nm$=8L_{1}/7$
which agrees with the setup in \citep{liu2007far}. We then placed
the charge closer to the top interface at $z=3L_{1}/4$
and $z=3L_{1}/8$.
\subsubsection{Charge located at $z_{0}=40\mathrm{nm}=8L_{1}/7$}
\begin{figure}[htb]
\caption{Real and imaginary parts of the potential for a charge located at
$z_{0}=40$nm$=8L_{1}/7$}
\label{Fig_potential_all_Regions_z_0_8_7_L1}
%\centering{}\includegraphics{\string"PMMA_silver_PR_ z_0=4.5714_4L_1/real_and_imag_Potential_all_Regions\string".eps}
%\centering{}\includegraphics[width=8cm]{real_and_imag_Potential_all_Regions-8_7_fixed}
\includegraphics[width=7.4cm]{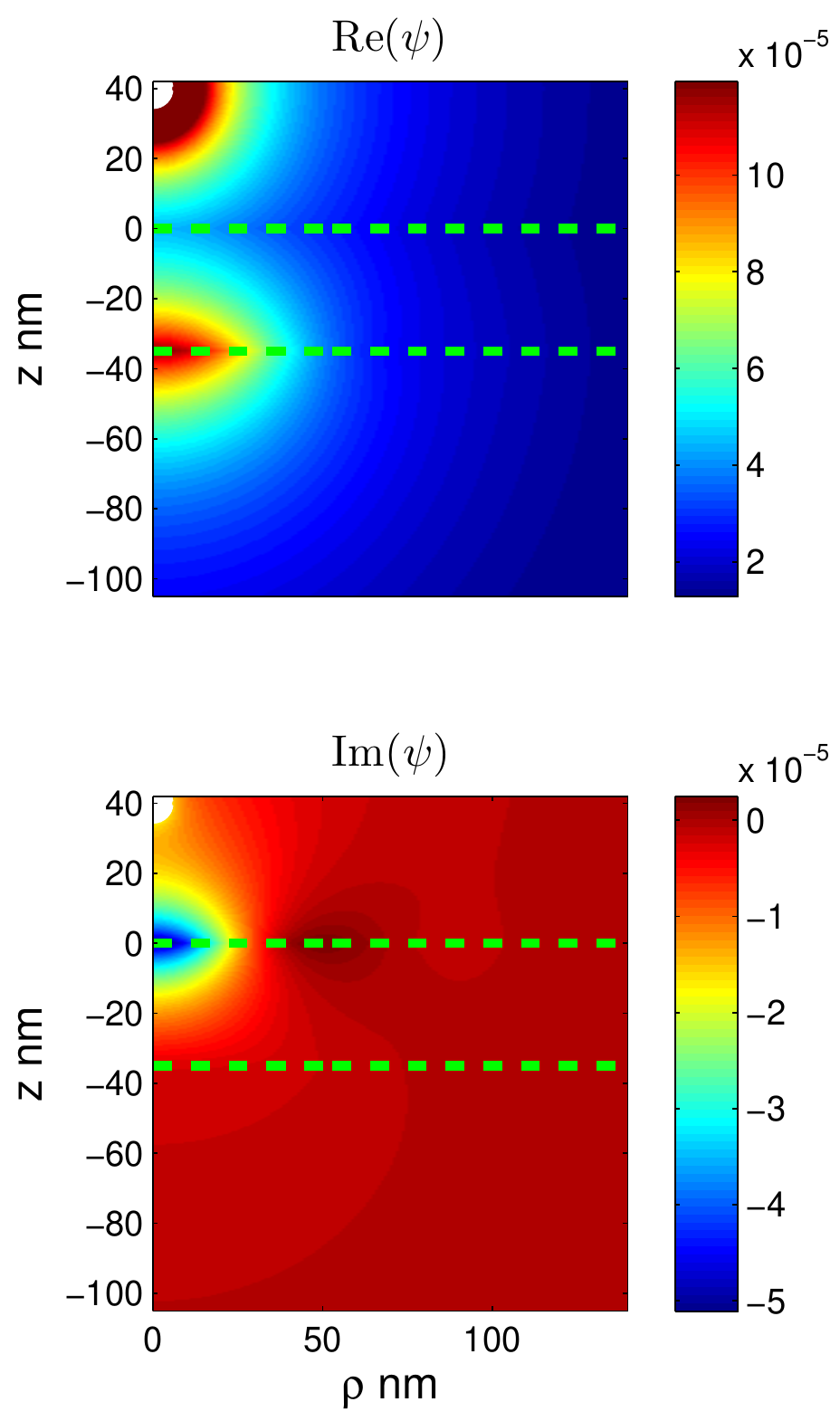}
\end{figure}

\begin{figure}[h]
\vspace{-2 true mm}
\caption{Intensity and dissipation for a charge located at $z_{0}=40$nm$=8L_{1}/7$}
\label{fig_intensity_dissipation_z0_8_7_L1}
%\centering{}\includegraphics[width=8cm]{\string"PMMA_silver_PR_z_0=4.5714_4L_1/Intensity_and_Dissipation_all_Regions\string".eps}
%\centering{}\includegraphics[width=8cm]{\string"PMMA_silver_PR_ z_0=3_4L_1/Intensity_and_Dissipation_all_Regions\string".eps}
%\centering{}\includegraphics[width=8cm]{Intensity_and_Dissipation_all_Regions-8_7_2}
\includegraphics[width=7.4cm]{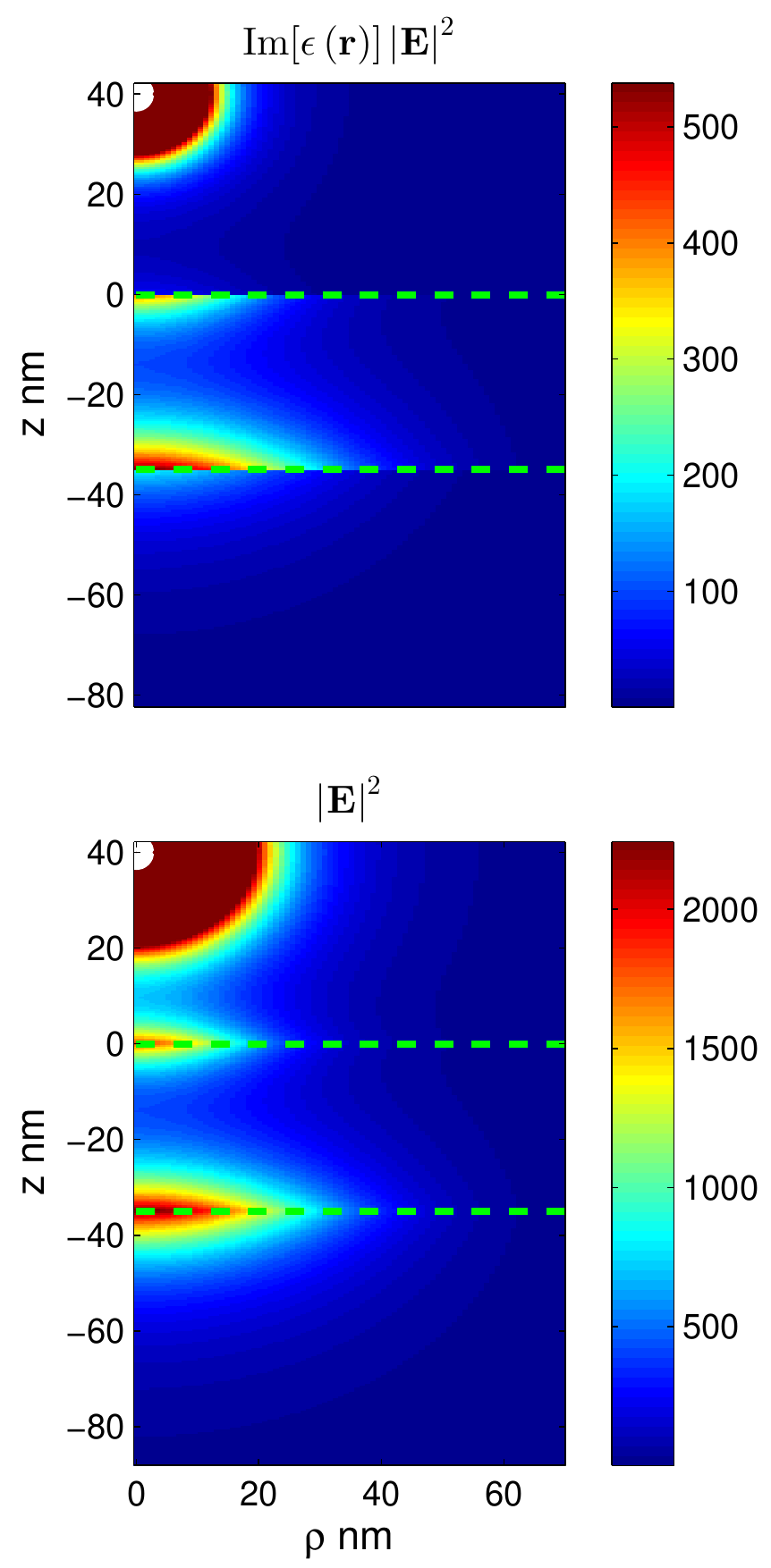}
\end{figure}

\begin{figure}[htb]
\caption{Intensity and horizontally normalized intensity in Region I for
two charges located at $z_{0}=40$nm$=8L_{1}/7$, $x_{1}=0,\, x_{2}=82.4$nm}
\label{fig_two_charges_z0_8_7_L1}
%\includegraphics{\string"PMMA_silver_PR_ z_0=4.5714_4L_1/Two_charges_Intensity_Reg_1\string".eps}
%\centering{}\includegraphics[width=8cm]{\string"PMMA_silver_PR_ z_0=4.5714_4L_1/Two_charges_Intensity_Reg_1\string".eps}
%\centering{}\includegraphics[width=7cm]{Two_charges_Intensity_Reg_1-eps-8_7}
\includegraphics[width=7.4cm]{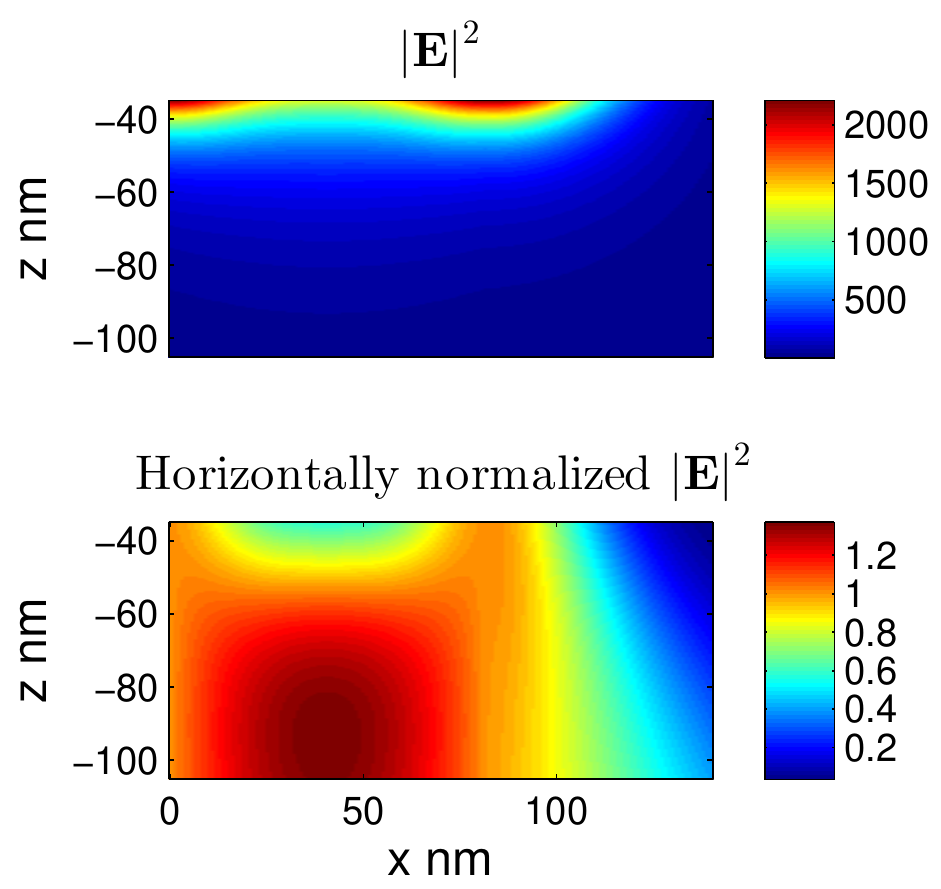}
\end{figure}

We first placed the charge  at $z_{0}=40$nm$=8L_{1}/7$ as in
\citep{liu2007far}. In Fig. \ref{Fig_potential_all_Regions_z_0_8_7_L1}
we present the real and imaginary parts of the potential in all the
regions. The potential is of course time dependent according to: 
$$\mathrm{Re}\left(\psi e^{i\omega t}\right)=\mathrm{Re}\left(\psi\right)\cos\left(\omega t\right)-\mathrm{Im}\left(\psi\right)\sin\left(\omega t\right). $$
The white circle denotes the object and, where applicable in the subsequent figures,
the image expected according to geometrical optics. 
It can be seen that Re$\left(\psi\right)$ has high values at
the $z=-L_{1}$ interface and that Im$\left(\psi\right)$
has high (absolute) values at the $z=0$ interface. In this figure,
as well as in subsequent figures that display all the regions,
we used a linear color scale. In order to present an
informative figure we mapped all the values higher than a certain value to this value. Thus, in all the locations which exhibit the highest value,
the actual values are often much higher than the apparent value.
In Fig. \ref{fig_intensity_dissipation_z0_8_7_L1} we present the
intensity and the dissipation in all the regions.
It can be seen that the intensity is high at the interfaces and has a
 higher value at the bottom interface. The dissipation in Region II
is higher than in Region I due to the fact
that the imaginary part of the permittivity is higher in Region II.
Note that the amplification of the electric field and the intensity starts even before
the top interface. This adds to the picture described in \citep{pendry2000negative}
where the amplification of the evanescent waves only in the silver slab was discussed. 
In Fig. \ref{fig_two_charges_z0_8_7_L1} we show the intensity
and the horizontally normalized intensity in Region I for two horizontally
displaced charges. It can be seen that the maximal resolution is at the interface $z=-L_{1}$.
The distance between the charges that enables the images to be resolved as previously explained  is $82.4$nm which is in good agreement with the results of \citep{liu2007far} - see Fig. 4D there.

\subsubsection{Charge located at $z_{0}=26.25\mathrm{nm}=3L_{1}/4$}
\begin{figure}[htb]
\caption{Real and imaginary parts of the potential for a charge located at
$z_{0}=26.25$nm$=3L_{1}/4$}
\label{real_imaginary_potential_z0_3_4_L1}
%\centering{}\includegraphics[width=8cm]{\string"PMMA_silver_PR_ z_0=3_4L_1/real_and_imag_Potential_all_regions\string".eps}
\centering{}\includegraphics[width=7.4cm]{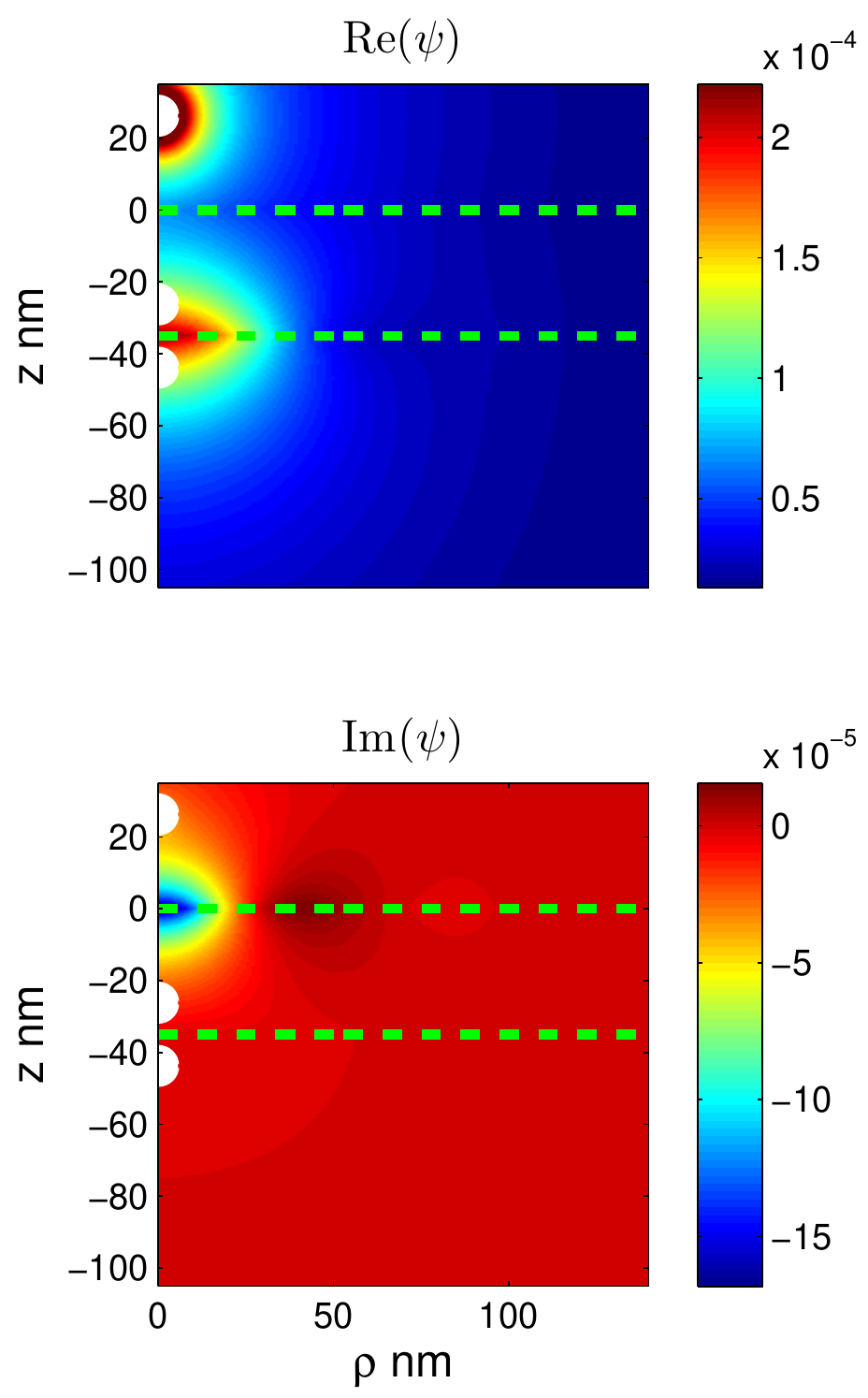}
\end{figure}

In Fig. \ref{real_imaginary_potential_z0_3_4_L1} we present the real
and imaginary parts of the potential in all regions for a charge located
at $z_{0}=26.25$nm$=3L_{1}/4$. 
Here too Re$\left(\psi\right)$ and Im$\left(\psi\right)$ (in absolute
value) peak at the bottom and top interface respectively.

\begin{figure}[htb]
\caption{Intensity and dissipation for a charge located at $z_{0}=26.25$nm$=3L_{1}/4$}
\label{fig_intensity_dissipation_z0_3_4_L1}
\centering{}\includegraphics[width=7.4cm]{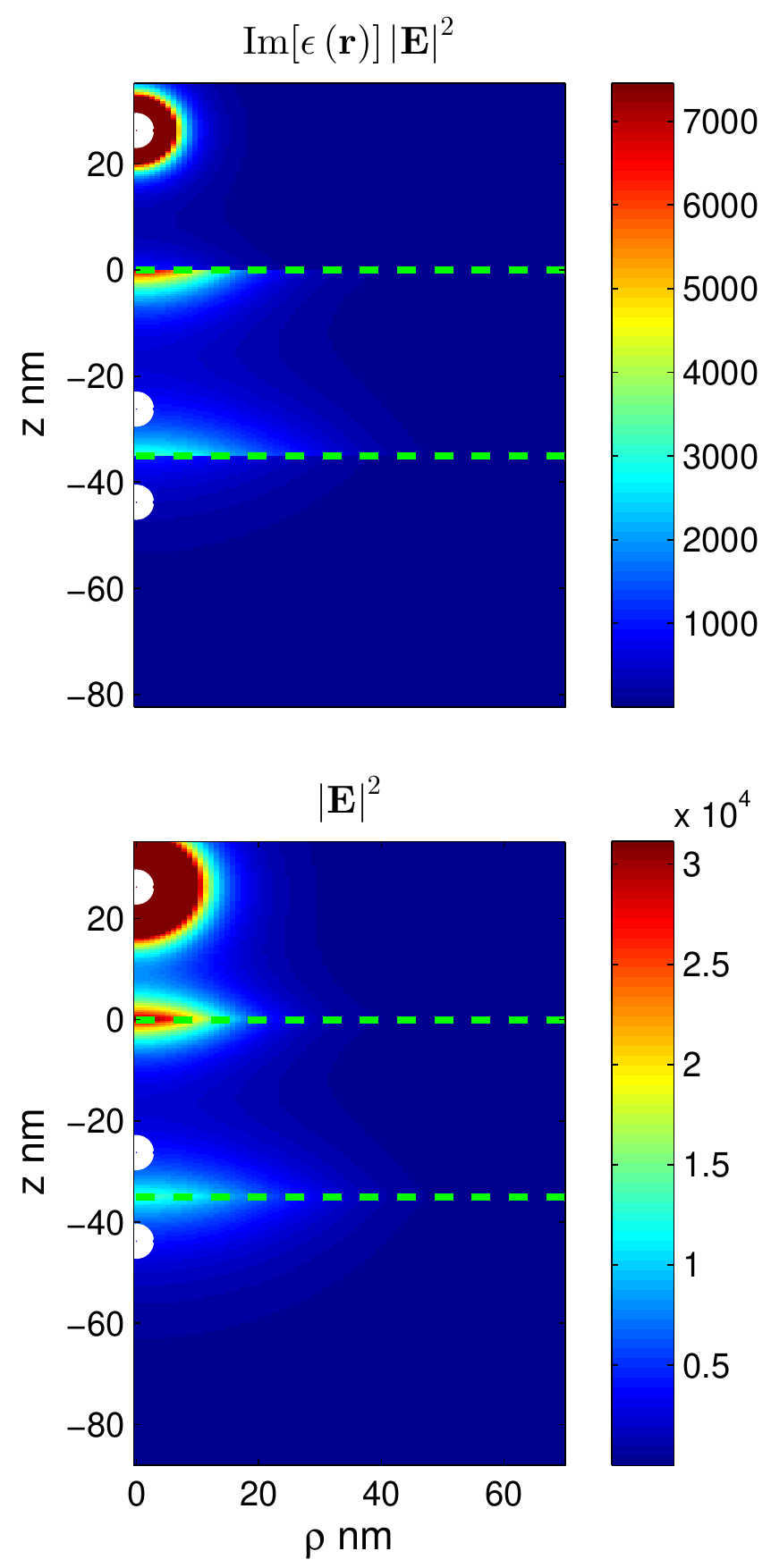}
\end{figure}

In Fig. \ref{fig_intensity_dissipation_z0_3_4_L1} we present the
intensity and the dissipation in all regions. 
Here we originally expected that the intensity would have high values at the geometric optics foci $z=-3L_{1}/4$ and at $z= z_{0}-2 L_{1}$. 
However, the intensity is actually concentrated
at the $z=0$ and $z=-L_{1}$ interfaces. Also, in this case 
the peak intensity is higher at the top interface.
The intensity in Region I is almost one order of magnitude higher than in the previous
case.

\begin{figure}[htb]
\caption{Intensity and horizontally normalized intensity in Region I for
two charges located at $z_{0}=26.25$nm$=3L_{1}/4$, $x_{1}=0,x_{2}=72$nm}
\label{fig_2_charges_z0_3_4_L1}
\begin{centering}
\includegraphics[width=7.4cm]{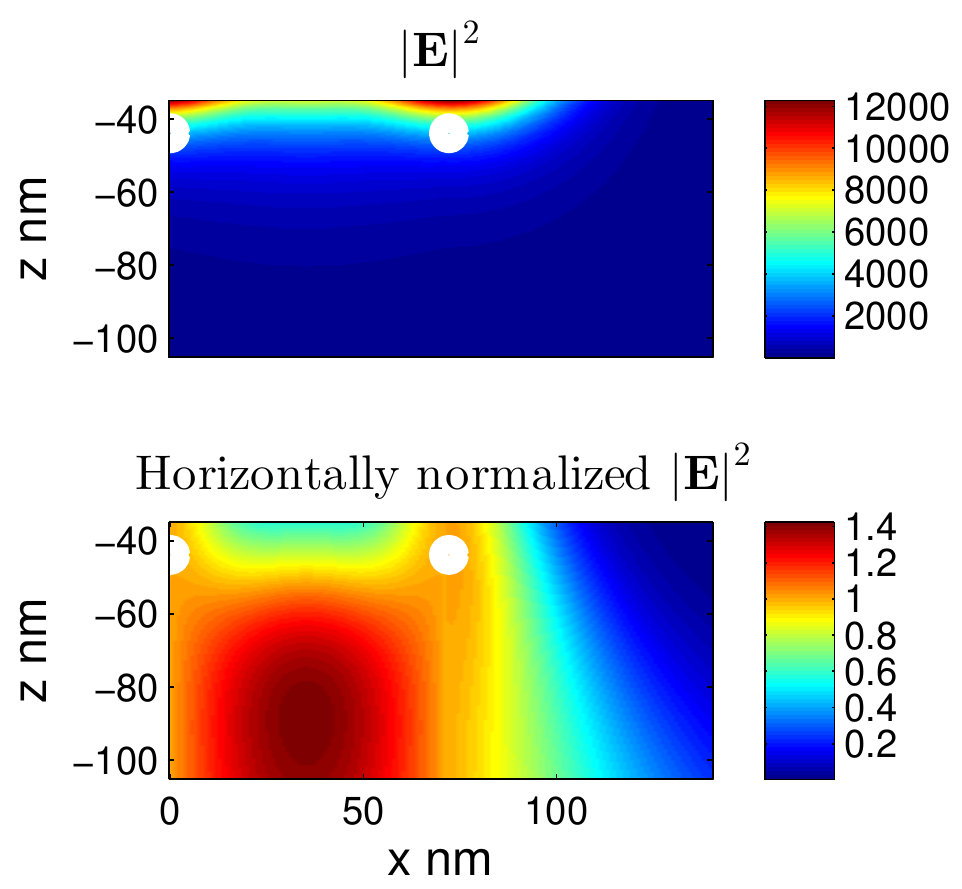}
\par\end{centering}
\end{figure}

In Fig. \ref{fig_2_charges_z0_3_4_L1} the intensity and the horizontally
normalized field intensity in Region I for two horizontally displaced charges are presented. The white circles denote the 
focal points. The separation exhibited is the smallest for which the images are still resolved as previously defined.
Surprisingly, the separation of the images is best not at the expected
focal plane but at the interface. Thus, both in terms of intensity
and resolution the image formed at the interface $z=-L_{1}$
is optimal.
In addition it can be seen that the separation distance in this case is $72\mathrm{nm}$
which is better than the former one.

\subsubsection{Charge located at $z_{0}=13.125\mathrm{nm}=3L_{1}/8$}
Here we calculate the potential, intensity and dissipation for a setup with a charge located at $z_{0}=3L_{1}/8$. 
In this case if  $s$ were equal to $1/2$ the dissipation rate should have diverged in the range $z_0-3L_1/2<z<-z_0-L_1/2$. However, since $s \neq 1/2$ 
we expect that the dissipation rate will increase in that range compared to the previous case where $z_{0}=3L_{1}/4$.    

\begin{figure}[htb]
\caption{Real and imaginary parts of the potential for a charge located at
$z_{0}=13.125$nm$=3L_{1}/8$}
\label{fig_real_imaginary_potential_z0_3_8_L1}
\centering{}\includegraphics[width=7.4cm]{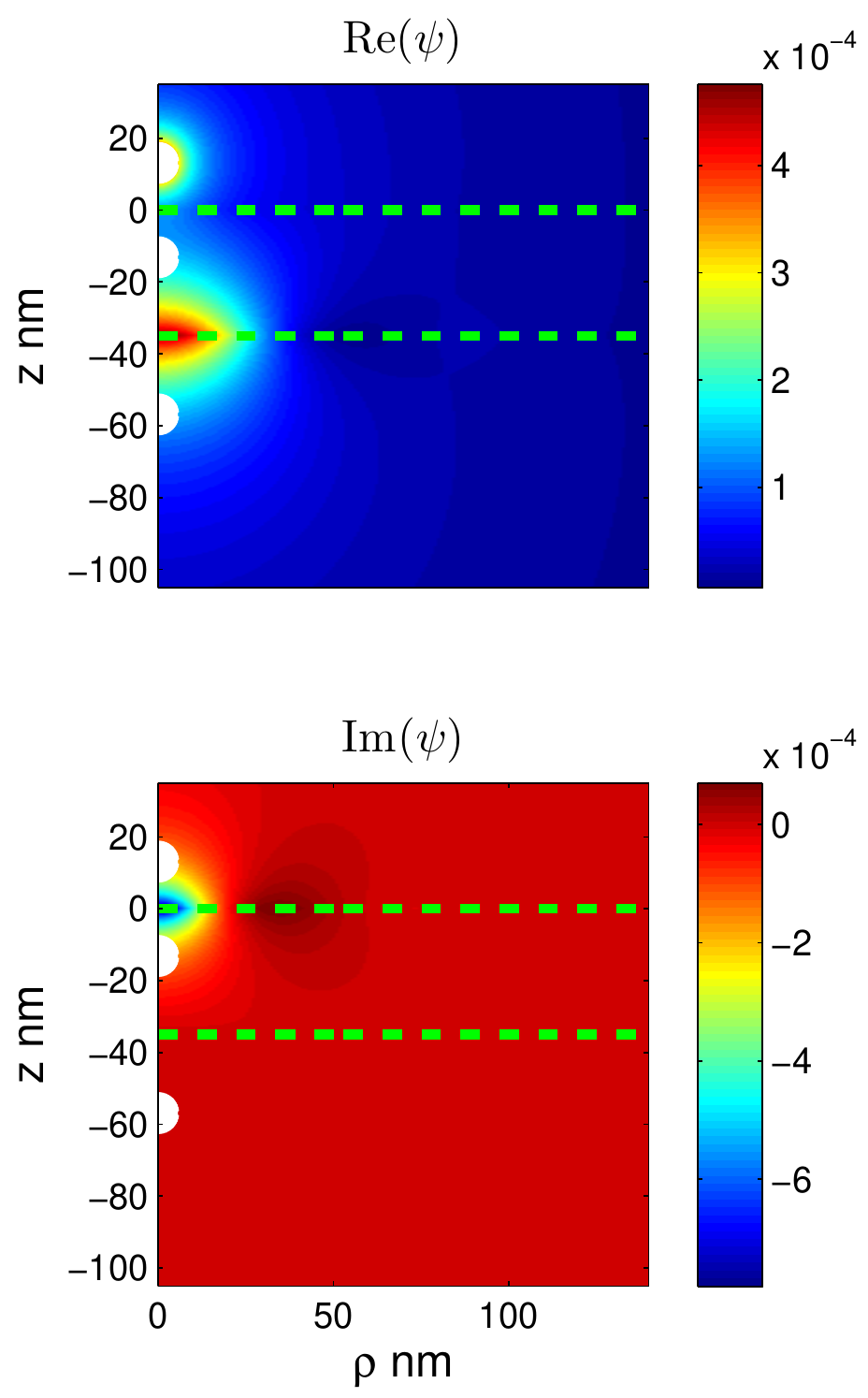}
\end{figure}

In Fig. \ref{fig_real_imaginary_potential_z0_3_8_L1} we present the
real and imaginary parts of the potential for a charge located at
$z_{0}=13.125$nm$=3L_{1}/8$. Here again Re$\left(\psi\right)$ and Im$\left(\psi\right)$ peak at the
bottom and top interfaces respectively.

\begin{figure}[htb]
\caption{Intensity and dissipation for a charge located at $z_{0}=13.125$nm$=3L_{1}/8$}
\label{fig_intensity_and_dissipation_z0_3_8_L1-2nd_scale}
\centering{}\includegraphics[width=7.4cm]{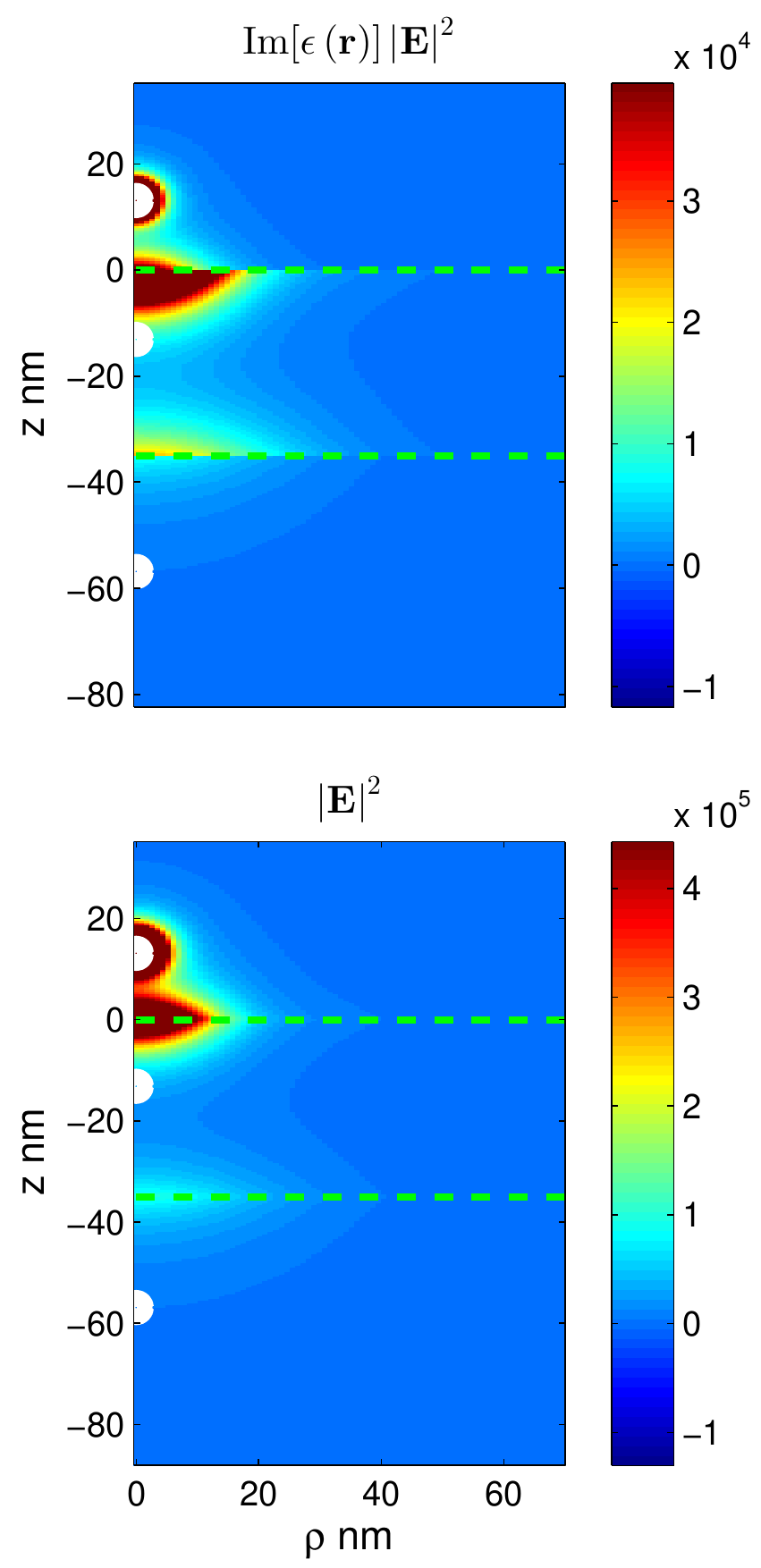}
\end{figure}

 In Fig. \ref{fig_intensity_and_dissipation_z0_3_8_L1-2nd_scale} we present
the intensity and dissipation in the 3 regions. Here too, the intensity is maximal at the interfaces rather than at the geometric optics foci.
It can be clearly seen that the intensity is higher at the top interface.
The intensity and the dissipation at the bottom interface in this case are almost one order of magnitude higher
than in the former case. 

\begin{figure}[htb]
\caption{Intensity and horizontally normalized intensity in Region I for
two charges located at $z_{0}=13.125$nm$=3L_{1}/8$,$x_{1}=0,x_{2}=63.2$nm}
\label{fig_2_charges_z0_3_8_L1}
\centering{}\includegraphics[width=7.4cm]{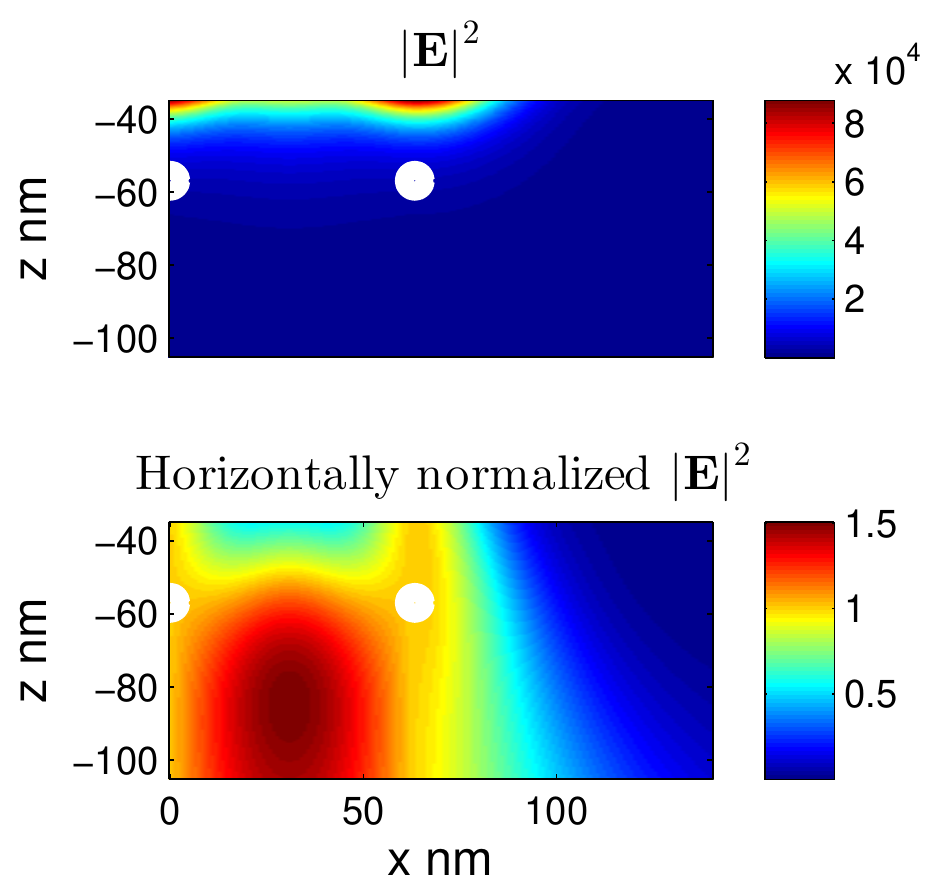}
\end{figure}

In Fig. \ref{fig_2_charges_z0_3_8_L1} we present the intensity and the horizontally normalized intensity
in Region I for two horizontally separated charges. It can be seen that the separation distance in this case is $63.2$nm
which is better than in the former cases.

It can be concluded that for the three object locations, the best images are formed
at the interfaces. As we moved the point charge closer to the $z=0$ interface,
the image formed at the $z=-L_{1}$ interface became better in terms of both intensity and resolution.

\subsection{Computations for other permittivities}
We repeated our computations for other values of $s$ for a charge
located at $z=3L_{1}/4$. First, we performed computations
with a setup in which the real part of $\Delta s$ remained the same
as in Subsection \ref{subs:different_vertical_charge_locations} and the imaginary part was divided by 100.
Then, we performed a computation in which both the real and imaginary
parts of $\Delta s$ were divided by 100.

\subsubsection{$\Delta s$ with $\mathrm{Im}\left(\Delta s\right)$ divided by 100}

\begin{figure}[htb]
\caption{Real and imaginary part of the potential $z_{0}=3L_{1}/4,\,\Delta s=0.0014+0.00032i$ }
\label{fig_real_imaginary_potential_z0_3_4_L1_imag_s_div_100}
\centering{}\includegraphics[width=7.4cm]{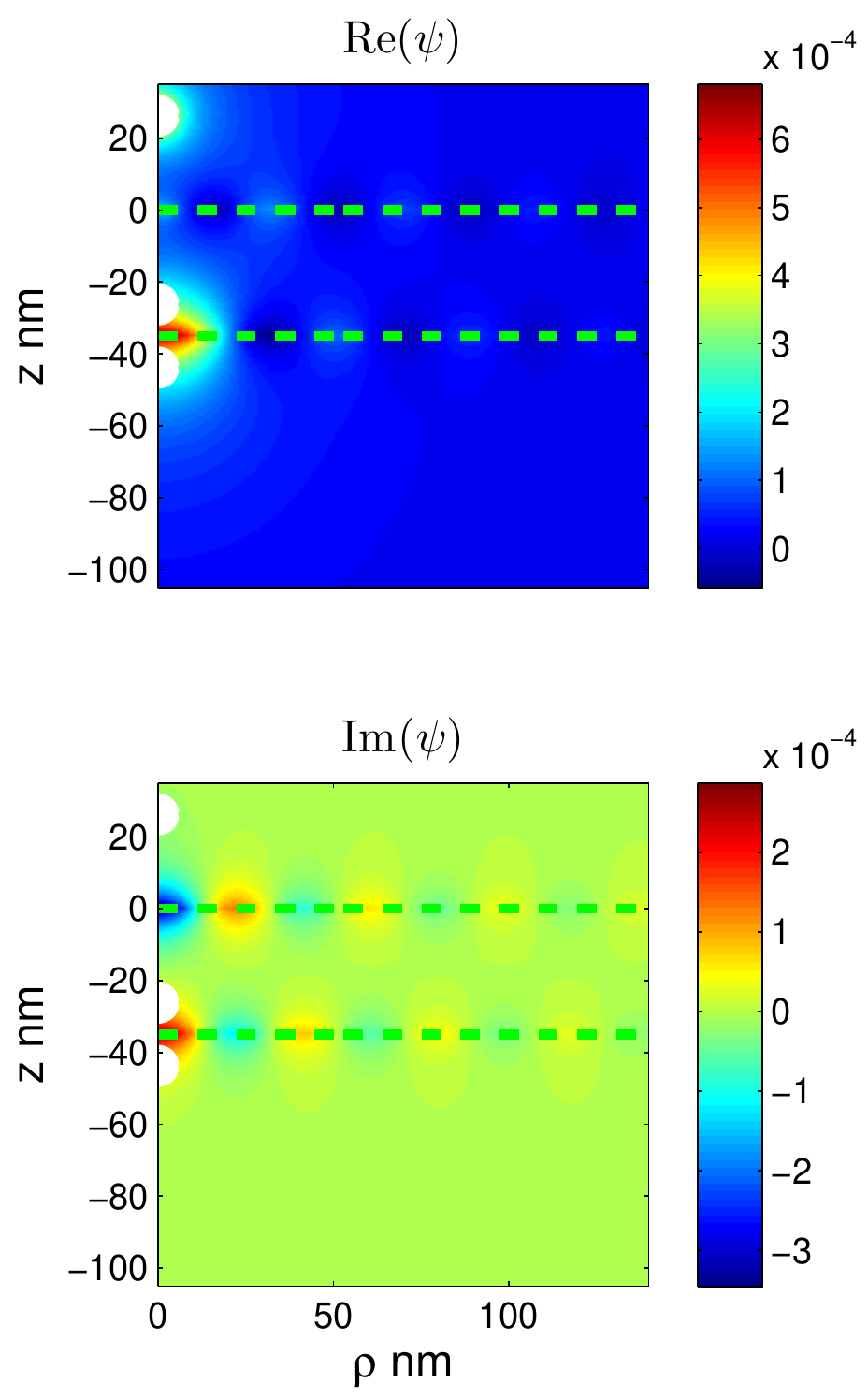}
\end{figure}

\begin{figure}[htb]
\caption{Intensity and dissipation for $z_{0}=3L_{1}/4\,\Delta s=0.0014+0.00032i$ }
\label{fig_intensity_dissipation_z0_3L1_4_imag_s_div_100}
\centering{}\includegraphics[width=7.4cm]{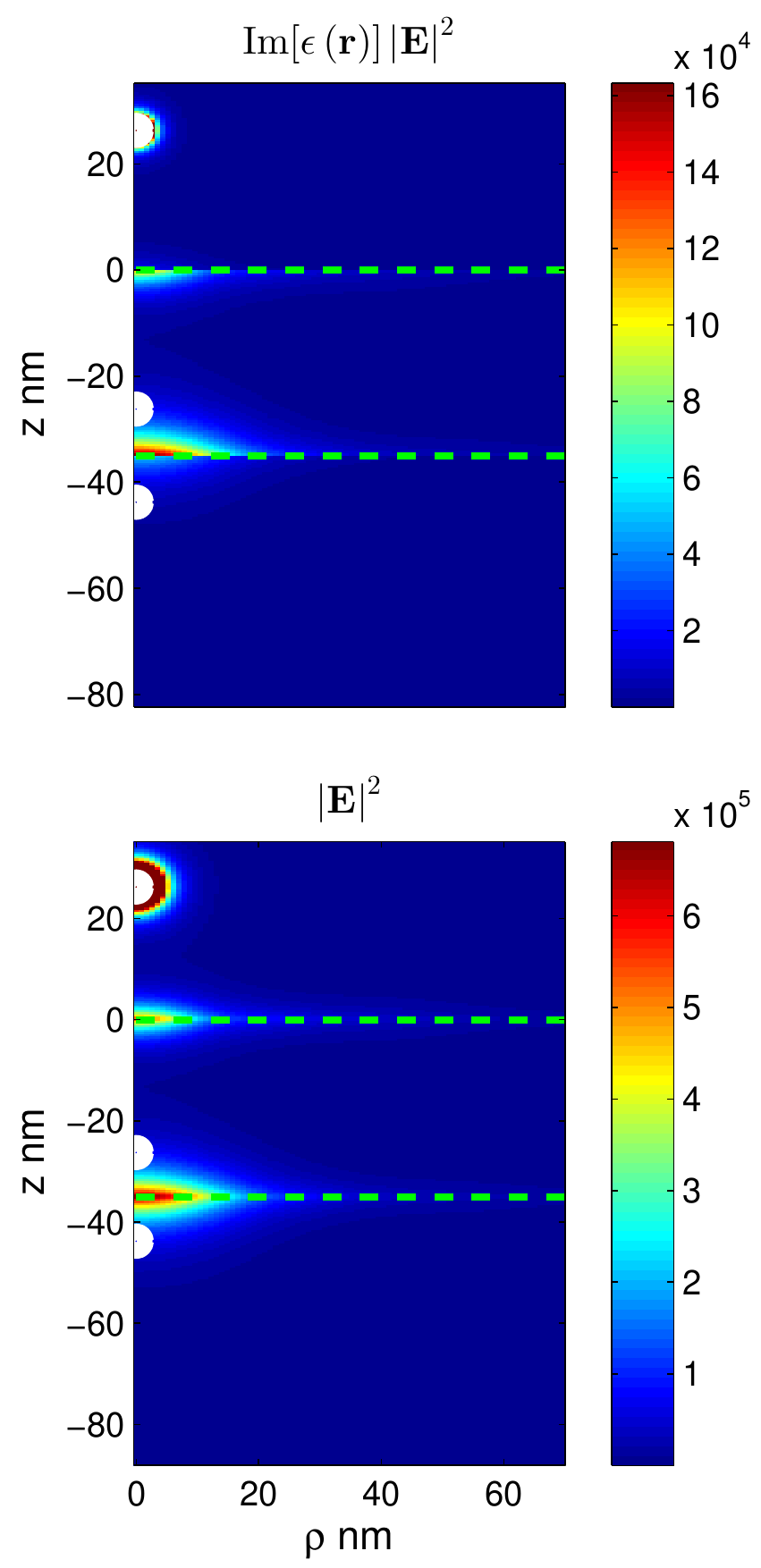}
\end{figure}

In Fig. \ref{fig_real_imaginary_potential_z0_3_4_L1_imag_s_div_100} Re$\left(\psi\right)$ and Im$\left(\psi\right)$ are presented. It can be seen that the potential now has 
an alternating sign as argued in \cite{Bergman2013static}.
In Fig. \ref{fig_intensity_dissipation_z0_3L1_4_imag_s_div_100} the
intensity and the dissipation are presented.
The intensity at both interfaces is higher than in the PMMA-silver-photoresist
setup. In addition the intensity here is higher at the bottom interface
as opposed to the previous setup with $z_{0}=3L_{1}/4$, where it was higher at the top
interface. The same is true regarding the local dissipation rates, despite the fact that Im$ \left(\Delta s\right)$ is smaller (which can be satisfied when the imaginary part of the permittivity is small everywhere in the system). This is due to the fact that $\psi$ and $\mathbf{E}$ tend to $\infty$ as $\Delta s\rightarrow0$ only at the lower interface.

\begin{figure}[htb]
\caption{Intensity and horizontally normalized intensity for 2 charges $z_{0}=3L_{1}/4,x_{1}=0,x_{2}=44.8$nm }
\label{fig_2_charges_z0_3_4_L1_imag_ds_div_100}
\centering{}\includegraphics[width=7.4cm]{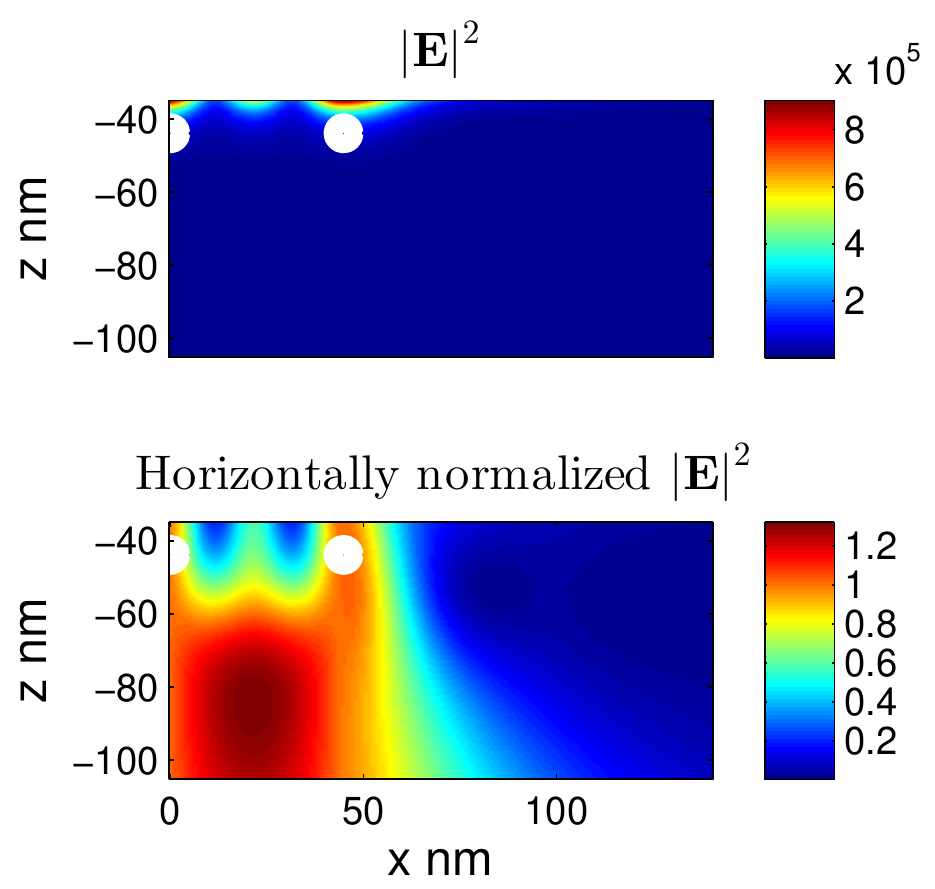}
\end{figure}

In Fig. \ref{fig_2_charges_z0_3_4_L1_imag_ds_div_100} the intensity
and the horizontally normalized intensity for 2 charges in Region
I are presented.
In this case the minimum separation distance between two objects for resolution of the images 
is $44.8$nm, which is significantly better than when Im$(\Delta s)$ was not decreased by a factor of 100.

\subsubsection{$\Delta s$ with both $\mathrm{Re}\left(\Delta s\right)$ and $\mathrm{Im}\left(\Delta s\right)$ divided by 100}
\begin{figure}[htb]
\caption{Real and imaginary part of the potential $z_{0}=3L_{1}/4,\,\Delta s=0.000014+0.00032i$ }
\label{fig_real_and_imaginary_potential_z0_3_4_L1_ds_real_imag_div_100}
\centering{}\includegraphics[width=7.4cm]{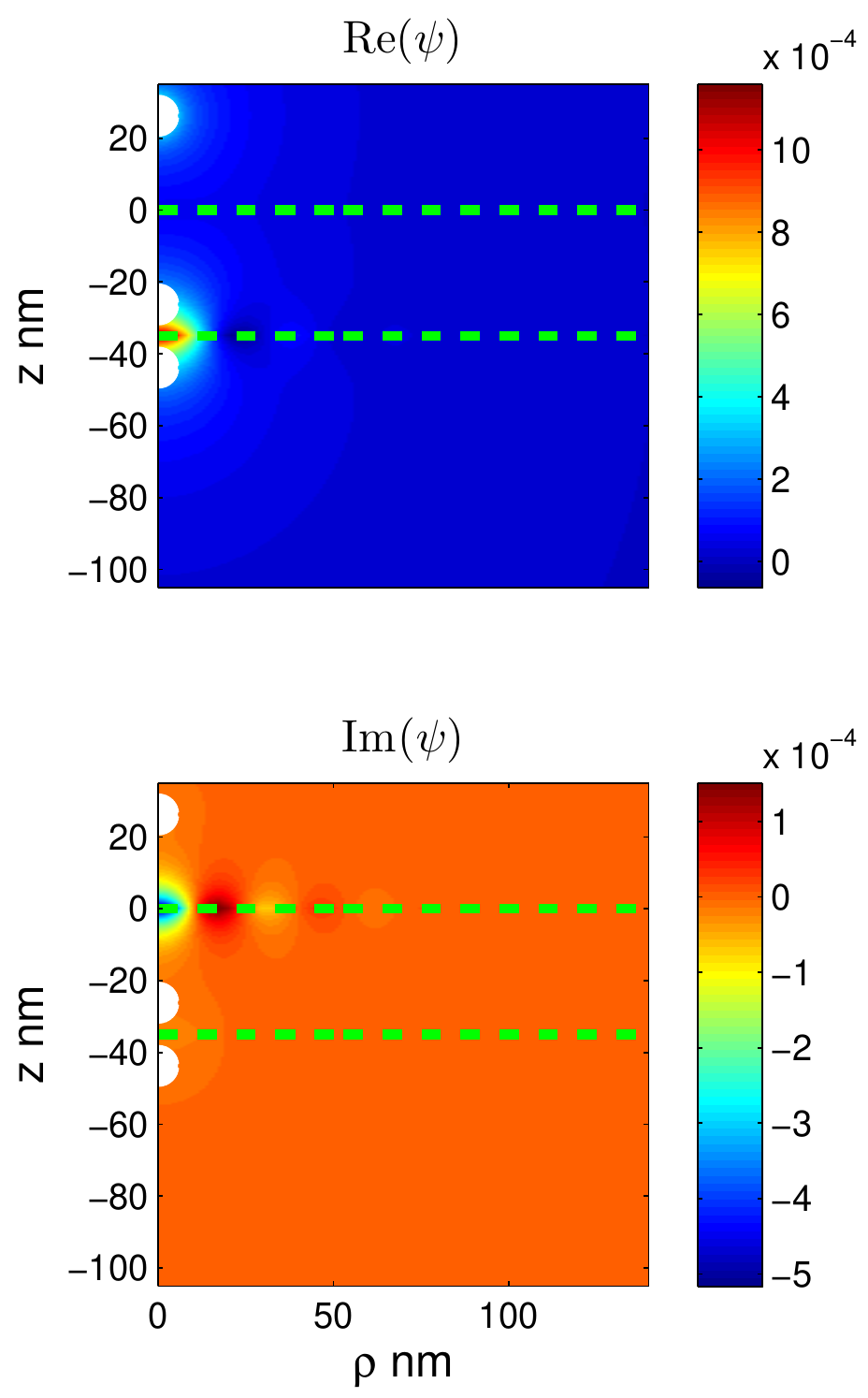}
\end{figure}

\begin{figure}[htb]
\caption{Intensity and dissipation for $z_{0}=3L_{1}/4,\,\Delta s=0.000014+0.00032i$}
\label{fig_intensity_dissipation_z0_3_4_L1_real_imag_ds_div_100}
\begin{centering}
\includegraphics[width=7.4cm]{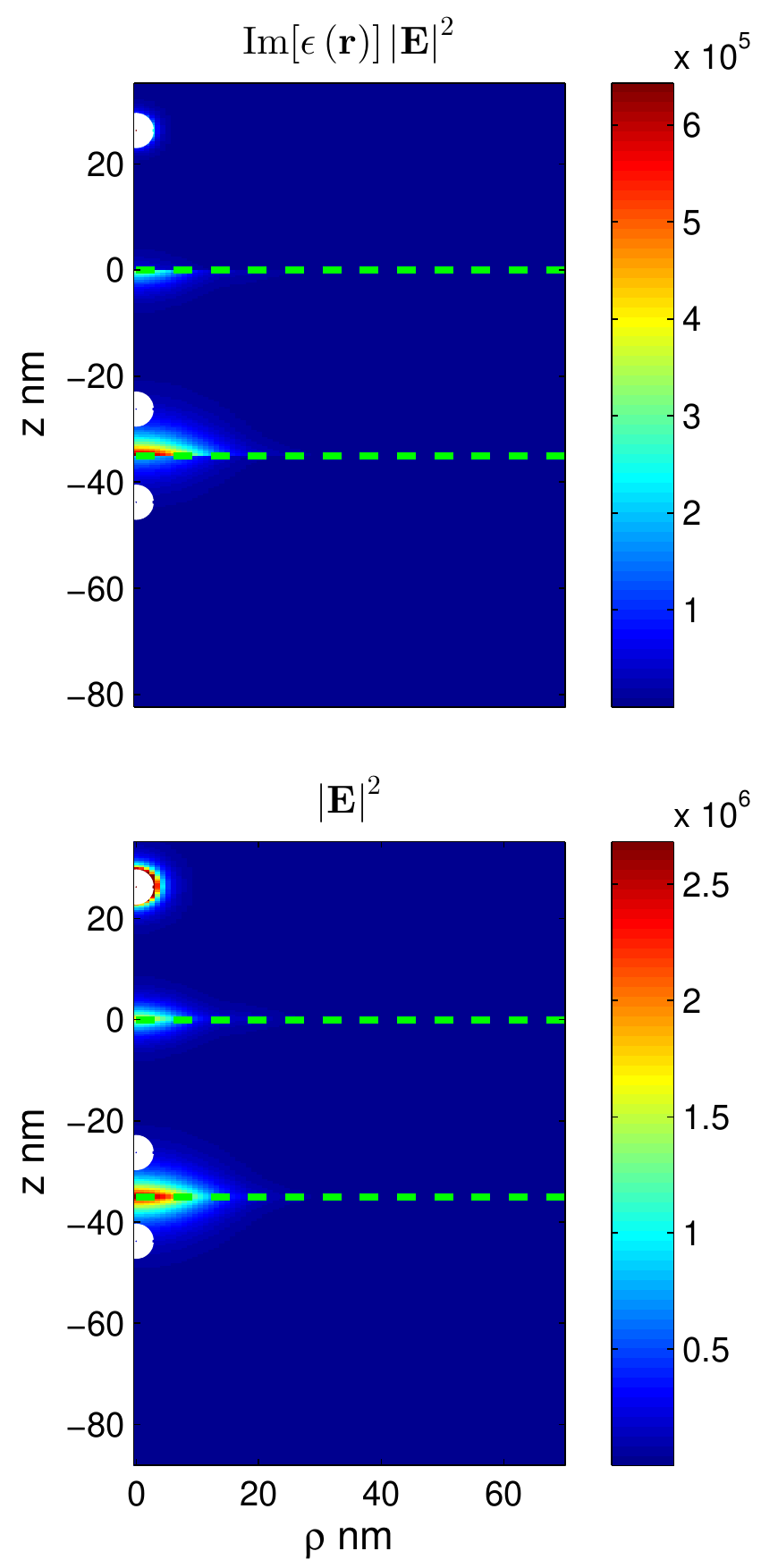}
\par\end{centering}
\end{figure}

\begin{figure}[htb]
\caption{Intensity and horizontally normalized intensity in Region I for
2 charges $z_{0}=3L_{1}/4,x_{1}=0,x_{2}=32$nm$\,,\Delta s=0.000014+0.00032i$}
\label{fig_2_charges_z0_3_4_L1_real_imag_ds_div_100}
\centering{}\includegraphics[width=7.4cm]{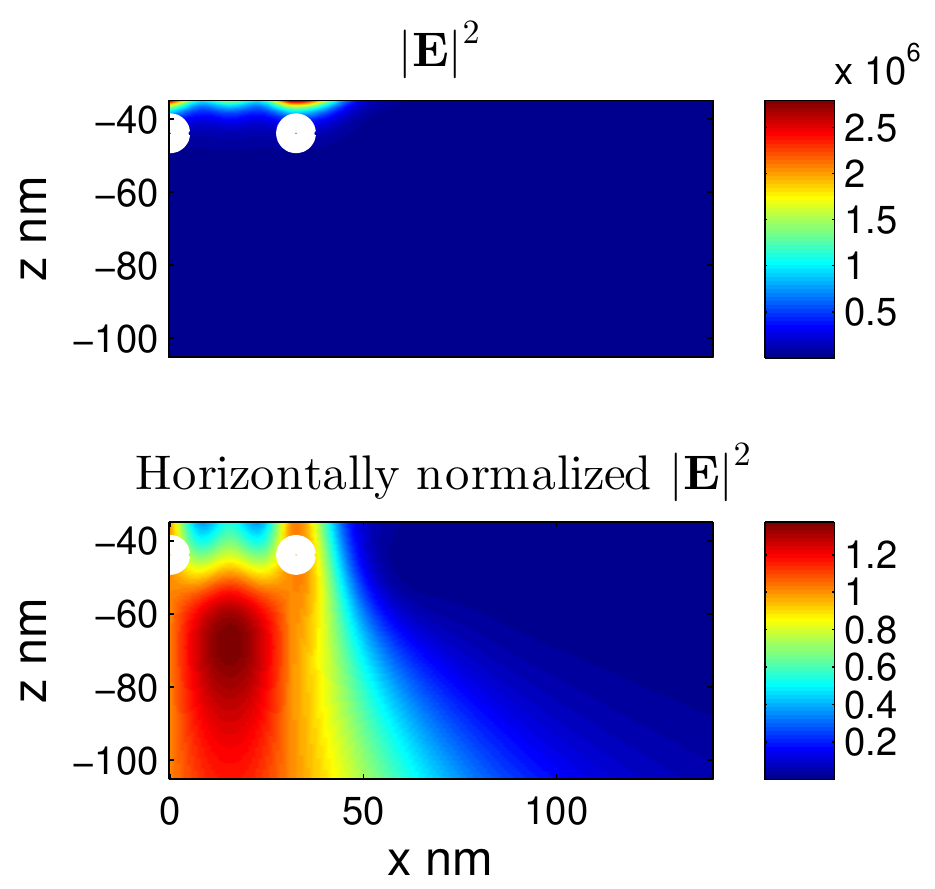}
\end{figure}

In Fig. \ref{fig_real_and_imaginary_potential_z0_3_4_L1_ds_real_imag_div_100}
Re$\left(\psi\right)$ and Im$\left(\psi\right)$ are presented. They peak (in absolute value) at the bottom and top interface respectively and they have alternating signs.
In Fig. \ref{fig_intensity_dissipation_z0_3_4_L1_real_imag_ds_div_100} the 
intensity and dissipation for all the regions are presented. It can be seen that $I$ and $W$ at the bottom interface are higher compared to the case
when we decreased just the imaginary part of $\Delta s$.
In Fig. \ref{fig_2_charges_z0_3_4_L1_real_imag_ds_div_100} the intensity
and the horizontally normalized intensity in Region I for two separated charge
objects are displayed.
The minimum separation distance for this $\Delta s$ value is $32$nm. It
can be seen that when we also decrease Re$\left(\Delta s\right)$
we have better separation between images.

We can conclude that when the value of $\Delta s$ is lowered, 
the optimal image locations are also at the interfaces. As we decrease
the real and imaginary parts of $\Delta s$ both the intensity and
the resolution become better for imaging. The analysis suggests
that in the quasistatic regime for a setup with a small value of $\Delta s$,
very high intensity and resolution can be reached (this occurs when $\epsilon_{1}\thickapprox-\epsilon_{2}$). It would be interesting to investigate whether such a pair
of materials exists or can be engineered.

\subsection{Analysis for a definite value of $k$}
The expressions for the potential and the electric field can be easily decomposed into their $k$ components. Namely,
the component associated with a specific $k=\left|\mathbf{k}\right|$ is simply the integrand
 in Eqs. \eqref{psi2}, \eqref{eq:er_reg_2}, \eqref{eq:e_z_reg_2}.
Thus, we can easily calculate the contribution of each $k$ component
to the potential and the electric field.

It was interesting, in the case where $\Delta s\rightarrow0$,
to calculate the amplitude of the  electric field for a given $\rho$ and $k$ values at the top interface and compare it to
the same quantity at the bottom interface. To that end we substituted $z=0$ and $z=-L_1$ in the integrands of Eqs. \eqref{eq:er_reg_2}, \eqref{eq:e_z_reg_2} and took the limit $\Delta s\rightarrow0$.
This leads to the following results:
\begin{eqnarray*}
\frac{\underset{\Delta s\rightarrow0}{lim}E_{\textrm{II}\, \rho}\left(z=-L_{1}\right)}{\underset{\Delta s\rightarrow0}{lim}E_{\textrm{II}\, \rho}\left(z=0\right)}=\frac{C_{2}kJ_{1}\left(k\rho\right)e^{-z_{0}k}e^{L_{1}k}}{C_{2}kJ_{1}\left(k\rho\right)e^{-z_{0}k}}=e^{L_{1}k},\\
\frac{\underset{\Delta s\rightarrow0}{lim}E_{\textrm{II}\, z}\left(z=-L_{1}\right)}{\underset{\Delta s\rightarrow0}{lim}E_{\textrm{II}\, z}\left(z=0\right)}=\frac{C_{2}kJ_{0}\left(k\rho\right)e^{-z_{0}k}e^{L_{1}k}}{C_{2}kJ_{0}\left(k\rho\right)e^{-z_{0}k}}=e^{L_{1}k}.
\end{eqnarray*}
These $k$ dependent ratios are the same as the transmission coefficient of the slab derived from the multiple scattering calculation in Ref. \citep{pendry2000negative}. This is another confirmation of the validity of our  results.
%It is worth noting that this ratio of the \emph{total} intensities associated with a given $k$ value at the interfaces is the same as the squared transmission coefficient of the slab for the same $k$ value derived in \citep{pendry2000negative}. This transmission %coefficient was calculated for a general plane wave solution of Maxwell's equations impinging on the top interface and then the $\omega\rightarrow0$ limit was taken. In the context of plane waves, here we effectively summed over all the ``plane waves''  %that reach the top interface and all the ``plane waves'' that leave the bottom interface. Thus, since in our calculation all the plane waves impinging on the top interface have the same $k$ value as the plane wave impinging on the top interface in %\citep{pendry2000negative}, we should get the same ratio.

\section{Discussion}
 \label{section:discussion}
We analyzed a two constituents setup of three dielectric slabs,  in which an electric point charge is located in the top slab.
We first derived exact expressions for the local electric field in the form of one dimensional integrals and verified our results.
We then performed numerical computations of the electric potential, intensity and dissipation for a setup that was previously tested in experiments.
We calculated these quantities of interest for several charge locations and several permittivity values. Finally we showed that our results agree with previous analytic results.

The computations reveal several important effects. The best images are formed at the interfaces between the slab and the surrounding medium rather than at the geometric optics foci.
This optimality is in terms of both intensity and resolution. In addition the computations confirm previous analysis in which it was stated that the dissipation rate diverges when $\epsilon_1=-\epsilon_2$. This can occur either when this quantity is real, in which case the constituents are free of any dissipation, or when they have imaginary parts with opposite signs. In the latter case one of the constituents exhibits dissipation while the other exhibits gain. As was explained in that analysis, this counterintuitive effect originates from the fact that $s=1/2$ is the accumulation point of all the eigenvalues and is therefore a very singular point of Maxwell's equations \cite{bergman2014perfect,Bergman2013static}.
The computations for several charge locations show that when the object is closer to the interface with the intermediate slab, the imaging is better. The computations for several permittivity values show that as $\epsilon_1 \rightarrow -\epsilon_2$
the imaging becomes better.  
\appendix
\section*{Appendix}
\section*{Verification of the analytic results}
\begin{widetext}
\begin{eqnarray}
\psi_{\textrm{I}}\left(z=-L_{1}\right)=\psi_{\textrm{II}}\left(z=-L_{1}\right)=\frac{4qs\left(1-s\right)}{\epsilon_{2}}\int e^{k\left(-L_{1}-z_{0}\right)}\left[\frac{1}{e^{-2kL_{1}}-\left(1-2s\right)^{2}}\right]J_{0}\left(k\rho\right)dk\\
%\end{equation}
%\begin{equation}
\psi_{\textrm{II}}\left(z=0\right)=\psi_{\textrm{III}}\left(z=0\right)=\frac{q\left(2\Delta s+1\right)}{\epsilon_{2}}\intop_{0}^{\infty}dkJ_{0}\left(k\rho\right)e^{-k\left(z_{0}\right)}\frac{e^{-2kL_{1}}-2\left(\Delta s\right)}{e^{-2kL_{1}}-4\left(\Delta s\right)^{2}}\\
%\end{equation}
%\begin{equation}
\epsilon_{1}E_{\textrm{II}_{z}}\left(z=-L_{1}\right)=\epsilon_{2}E_{\textrm{I}_{z}}\left(z=-L_{1}\right)=\epsilon_{1}\frac{4q\left(\frac{1}{2}+\Delta s\right)^{2}}{\epsilon_{2}}\intop_{0}^{\infty}dkkJ_{0}\left(k\rho\right)e^{k\left(-L_{1}-z_{0}\right)}\frac{1}{e^{-2kL_{1}}-4\left(\Delta s\right)^{2}}\\
%\end{equation}
%\begin{equation}
\epsilon_{1}E_{\textrm{II}_{z}}\left(z=0\right)=\epsilon_{2}E_{III_{z}}\left(z=0\right)=\epsilon_{1}\frac{\left(1+2\Delta s\right)q}{\epsilon_{2}}\intop_{0}^{\infty}dkkJ_{0}\left(k\rho\right)e^{-kz_{0}}\frac{e^{-2kL_{1}}+2\Delta s}{e^{-2kL_{1}}-4\left(\Delta s\right)^{2}}
\end{eqnarray}
where we used $\frac{\epsilon_{1}}{\epsilon_{2}-\epsilon_{1}}=\frac{\epsilon_{1}}{\epsilon_{2}}\left(\frac{1}{2}+\Delta s\right)=-\left(\frac{1}{2}-\Delta s\right)$
from the definition of $s$.
\end{widetext}

\end{document}